\documentstyle[12pt]{article}
\begin{document}

\begin{titlepage}
\begin{center}
{\bf Ability to Count to Two. Opening Talk at the
Third International Sakharov Conference on Physics}
\footnote{Moscow, Lebedev Institute, June 24-29, 2002;
www.sakharov.lpi.ru}

\vspace{1cm}

Boris L. Altshuler \footnote{E-mail:
altshuler@mtu-net.ru \& altshul@lpi.ru}

{\it Theoretical Physics Department, P.N. Lebedev Physical Institute,
53 Leninski Prospect, Moscow, 117991, Russia}
\end{center}

\vspace{1cm}

\begin{abstract}
Third International Sakharov Conference on Physics
organized by the Theoretical Physics Department of
the Lebedev Institute, Russian Academy of Sciences,
covered wide scope of topics: astrophysics, fusion,
high field, high pressure and high density research,
superstrings and dualities, brane world and quantum
gravity, quantum field theory and high energy physics.
This short Opening Word however exceeds the bounds
of physics - it is about three "dynamical characteristics"
of Sakharov's methodology and way of thinking: "Artseulov
method", "a permanent feeling of possible personal error",
and ability "to count to two", which in present days is
not less demanded than before.
\end{abstract}
\end{titlepage}

\vspace{5mm}

It is a pleasure and an honor to welcome participants
and guests of the Third International Sakharov Conference
on Physics here in the Lebedev Physical Institute. There
is a number of intimate connections. Dmitry Sakharov,
father of Andrei Sakharov, was a student of Petr Lebedev
in the Physical Department of Moscow University. Pressure
of light first observed by Lebedev was 40 years later
used by Sakharov to trigger the H-bomb. There are also
moral parallels: Petr Lebedev, who was more than far from
politics, being a personality of high ethic principles
could not help resigning from the University in a protest
against Tsar Government repressions towards his colleagues.
His Lab was destroyed, and although private sponsors began
to build for him "Lebedev institute" his heart did not stand
the disaster, he died in March 1912 in the age 46, and
supposedly because of it did not receive Nobel Prize on
Physics he was nominated to for his discovery.

Actually this Conference should have taken place a year ago,
being dedicated to 80 years of Sakharov who was born in
1921. The reason for this delay was serious: in May 1999
Professor Efim Fradkin died, and in June 2000 Theoretical
Physics Department organized in his memory the big
international conference "Quantization, gauge fields and
strings". To organize within a year one more huge conference was
impossible. There is no doubt that Andrei Dmitrievich would
approve of our actions in the memory of Efim S. Fradkin,
whom he highly respected as personality and as scientist. He
also very much respected Professor David Kirzhnits who died
recently. It was David Kirzhnits who in his memoirs  about
Sakharov (see in~\cite{Kirzh}) most transparently compared
Sakharov's methodology with the feat of Russian pilot Konstantin
Artseulov during the First World War, who the first time in
history of aviation deliberately dropped down his airplane
into the mortal spin (thus committing a suicide in public,
according to the hundreds of observers' opinion) and safely
went out of it, creating the method which saved lives of
hundreds of pilots. You see: it is always quite interesting
to speak about Sakharov whose ideas and actions were really
very often considered as "mortal", but they proved to be
salutary for all of us after all.

In this short Opening Word I'll concentrate upon three "dynamical
characteristics" of Sakharov's "method" and his way of
thinking. {\bf The first one}, which may be called, {\bf "Artseulov method"}
I already described above with the help of David Kirzhnits's
metaphor: after Sakharov came to certain conclusion he preferred
to act most resolutely and also as a good teacher. But to come
to the conclusion was not a simple dynamical process. {\bf Second
feature} of Sakharov's mentality may be called {\bf "a permanent
feeling of possible personal error"}. He writes in "Memoirs":

"My statements on general issues are often tentative, meant to
provoke discussion, and subject to revision. I agree with
Leszek Kolakowski when he writes: {\it "Inconsistency is simply
a secret awareness of the contradictions of this world... a
permanent feeling of possible personal error, or if not
that, then of the possibility that one's antagonist is right"}.
My only quarrel, - comments Sakharov after this quotation of
Kulakowski, - is with the word "inconsistency", which I would
replace with one that conveys my belief that intellectual
growth and social awareness should combine dynamic self-criticism
and a set of stable values... I am not a professional
politician. Perhaps that is why I am always burdened by
doubt about the usefulness and consequences of my
actions. I inclined to the belief that a combination of
moral criteria and unrestricted inquiry provides the
only possible compass" (\cite{Sakh}, pp. 579-580).

It is difficult to count how many times these Sakharov's
sincere words "I am not a professional politician..." were
used against him by Soviet propaganda. But  we are
talking now about something much more serious.

Sakharov really was a "thinking matter", everybody who
personally knew him, I also was happy to contact
him during 20 years, will confirm that his brain worked
permanently and any moment he was ready to contemplate
the problems of interest "as if he had a blank sheet
of paper in front of him", - as Igor Evgenievich
Tamm, his teacher and many years Chief of the Theoretical
Physics Department, put it. The same was true with
regards to Sakharov's own opinions and views: that is
why it was so interesting to speak to him. But then
the important question arises: If there were permanent
doubts and "unrestricted inquiry" (in Sakharov's words
cited above) how he managed to come to definite
conclusions, why he was not a sort of "Chekhov's hero"
incapable of any decision-making? Sakharov definitely
was not a "Chekhov's hero", but his decision-makings
were often rather difficult. It is not by chance
that in quotations above he speaks about "stable values"
and "moral criteria", i.e. about invariants which may
serve a certain guideline. Idea of  absolute priority
of individual human rights - above any social, national,
religious and other so called "great ideas" - was a
product of permanent internal search of such an invariants.

But to explain why he reached so much it is also
necessary to point out {\bf one more feature of
his mentality} - his special talent
which is very much advisable although not easy
to learn from him: he was good at high
mathematics, he was able, so to say, {\bf "to count to two"}.
The holism (in the Oriental language) of his thinking
is well known. He could consider the problem at hand
all at once, in all its complexity, with all its
particulars, and, this being most important, in
its dynamics. A space-time picture was obtained right
away, with expected answer in the end. And all
this, including actions already fulfilled or planned
ones, was contemplated over again and again. Why I
call this ability "to count to two"?  During half of
XX-th century half of Mankind admired Soviet socialism
with its ideas of social justice (ONE) being incapable
of noticing terror, GULAG etc. (TWO). Sakharov's idea
of convergence, although being questionable by
itself, was a sigh of relief for many and many.

One more example: "peacemaking" by western
intellectual elite (there is no sense to argue against
the good goal - peace is better than war, it is highest
humane priority - ONE) in practice unfortunately was
helping to push the Mankind closer to the
thermonuclear abyss because "elite" did not manage
to notice another side of the medal - aggressiveness
of militarized Soviet closed society and hence necessity
for the West (just for the sake of keeping and
strengthening Peace) to be sufficiently strong (TWO).
It was Sakharov who had a courage to reveal the truth
speaking from inside the USSR, who supported "double
solution" of NATO, which included additional deployment
of rockets with nuclear warheads targeted at the Soviet
cities (isn't it the Artseulov method?). Of course
because of it he himself became a target of severe
revenge of Soviet authorities; then "elite", colleagues
supported Sakharov. As a result of this dialectic
"Perestroika" began miraculously, Ronald Reagan came
to the Red Square and rockets with nuclear charges
began to be destroyed after all. (This time Mankind
avoided the fall into the abyss, although it was
real narrow escape; Sakharov was quite aware of the
unacceptably high probability of the "fall").

Ability
to see simultaneously both sides of a medal, to
possess this "mental mirrors" is really a rear talent.
This Sakharov's talent "to count to two" now is not
less demanded than before. But some of you in this Hall
may ask why I cover this moral and political issues
at the physical conference? The answer is two-fold:

(1) Because this is "Sakharov" conference. Harry
Lipkin from the Weizmann Institute of Science
writes in \cite{Lipk}: "Andrei Dmitrievich had the
remarkable ability to understand how systems
function - all systems, social, political, scientific,
technological, as well as interfaces between them. In
a sense he was a kind of an interdisciplinary systems
super-engineer". That is why it seems reasonable even
at this conference to exceed limits of physics, at
least in short.

(2) Because we all know the value of speed of light, but
what is the value of "speed of darkness"?  Will the Mankind
manage to win the race with mortal dangers capable
of "putting off the Sun" for our civilization?

Sakharov died almost 13 years ago. Time moves on and
now there are new hopes and new serious threats. World
terror with its new weapon - suicide bombers, who
potentially may be carriers of biological or nuclear
tools of mass destruction. In this connection I
draw attention to the most important and well balanced
Statement by Elena Bonner (April 2002) where she
outlined the global danger of this new weapon. Again
plenty of people prove to be incapable "to count to
two", got lost in two simple notions: "struggle with
terror" and "human rights". Those who advocate active
struggle with world terror sometimes have evident
difficulties with embracing demands of human rights
and humanity toward peaceful population - in this way
deliberately or undeliberately supporting terrorists
(unceasing  drama in Chechnya is a dreadful example).
Others - like now in Europe - seem to advocate human
rights but ignore the vital necessity of taking active
measures against terrorists - in this way also actively
supporting terrorists and hence most terrible violations
of human rights. I feel obliged to tell that members of
the Theoretical Physics Department joined the collective
protest of scientists against Israel-baiting campaign in
Europe. This campaign visually shows that we are still
too close to the precipice, and that Sakharov's talent
to comprehend two sides of a medal simultaneously is
strongly demanded, as well as his understanding that
individual human rights must be considered above any
national, religious and other corporate goals.

There are another dangers. General admiration of
achievements in biophysics and gene engineering is
accompanied by concerns about possibility of creating
artificial virus capable to kill the civilization on
the planet of Earth (is not our life "a permanent narrow
escape"?). Steven Hawking recently said in an interview
that he considers the death of Mankind because of such
viruses inevitable and the only way out is our settling
in cosmos. In this connection I would like to suggest
perhaps even more secure option of settling the Mankind
in the extra dimensions; this may be discussed at the
Conference. In physics, like in the life of human society,
modern days seem to be also boiling and interesting. Of
course Sakharov would be happy to participate in this
Conference. And I wish it success in its work.

\end{document}